\input harvmac
\input amssym.tex
\input amssym.def

\font\bbold=bbm10
\def\mC{\hbox{\bbold C}}

\def\mCP{\hbox{\bbold CP}}
\def\mWCP{\hbox{\bbold WCP}}

\def\mPT{\hbox{\bbold PT}}

\noblackbox
\newcount\footnotecount
\def\fn#1{\global\advance\footnotecount by
\footnote{$^{\the\footnotecount}$}{#1}}
\newcount\figno
\figno=0
\def\fig#1#2#3{
\par\begingroup\parindent=0pt\leftskip=1cm\rightskip=1cm\parindent=0pt
\baselineskip=11pt
\global\advance\figno by 1
\midinsert
\epsfxsize=#3
\centerline{\epsfbox{#2}}
\vskip 12pt
\centerline{{\bf Figure \the\figno} #1}\par
\endinsert\endgroup\par}
\def\figlabel#1{\xdef#1{\the\figno}}

\def\pmb#1{\setbox0=\hbox{#1}%
\kern-.025em\copy0\kern-\wd0
\kern.05em\copy0\kern-\wd0
\kern-.025em\raise.0433em\box0 }
\font\cmss=cmss10
\font\cmsss=cmss10 at 7pt
\def\inbar{\,\vrule height1.5ex width.4pt depth0pt}

\def\rlx{\relax\leavevmode}
\def\Cop{\relax\,\hbox{$\inbar\kern-.3em{\rm C}$}}
\def\Rop{\relax{\rm I\kern-.18em R}}
\def\Nop{\relax{\rm I\kern-.18em N}}
\def\one{\relax{\rm 1\kern-.25em I}}
\def\Pop{\relax{\rm I\kern-.18em P}}
\def\Zop{\rlx\leavevmode\ifmmode\mathchoice{\hbox{\cmss Z\kern-.4em Z}}
{\hbox{\cmss Z\kern-.4em Z}}{\lower.9pt\hbox{\cmsss Z\kern-.36em Z}}
{\lower1.2pt\hbox{\cmsss Z\kern-.36em Z}}\else{\cmss Z\kern-.4em
Z}\fi}


\def\ie{{\it i.e.}}


\def\sqr#1#2{{\vcenter{\vbox{\hrule height.#2pt
\hbox{\vrule width.#2pt height#1pt \kern#1pt
\vrule width.#2pt}\hrule height.#2pt}}}}


\def\makeblankbox#1#2{\hbox{\lower\dp0\vbox{\hidehrule{#1}{#2}%
   \kern -#1
   \hbox to \wd0{\hidevrule{#1}{#2}%
      \raise\ht0\vbox to #1{}
      \lower\dp0\vtop to #1{}
      \hfil\hidevrule{#2}{#1}}%
   \kern-#1\hidehrule{#2}{#1}}}%
}%
\def\hidehrule#1#2{\kern-#1\hrule height#1 depth#2 \kern-#2}%
\def\hidevrule#1#2{\kern-#1{\dimen0=#1\advance\dimen0 by #2\vrule
    width\dimen0}\kern-#2}%
\def\openbox{\ht0=1.2mm \dp0=1.2mm \wd0=2.4mm  \raise 2.75pt
\makeblankbox {.25pt} {.25pt}  }

\def\bun#1/#2{\leavevmode
   \kern.1em \raise .5ex \hbox{\the\scriptfont0 #1}%
   \kern-.1em $/$%
   \kern-.15em \lower .25ex \hbox{\the\scriptfont0 #2}%
}

\def\opensquare{\ht0=3.4mm \dp0=3.4mm \wd0=6.8mm  \raise 2.7pt
\makeblankbox {.25pt} {.25pt}  }


\def\sector#1#2{\ {\scriptstyle #1}\hskip 1mm
\mathop{\opensquare}\limits_{\lower 1mm\hbox{$\scriptstyle#2$}}\hskip
   1mm}

\def\tsector#1#2{\ {\scriptstyle #1}\hskip 1mm
\mathop{\opensquare}\limits_{\lower
   1mm\hbox{$\scriptstyle#2$}}^\sim\hskip  1mm}
\Title{\vbox{\baselineskip12pt
\hbox{DAMTP-2004-49}\hbox{SWAT-397}}}
{\vbox{\centerline{Strings in Twistor Superspace and Mirror Symmetry}
}}
\smallskip
\medskip\centerline{{\bf {S. Prem
Kumar${}^{a,b}$}\foot{S.P.Kumar@damtp.cam.ac.uk} and 
\bf{Giuseppe Policastro${}^{a}$}\foot{G.Policastro@damtp.cam.ac.uk}}}

\medskip
\centerline{${}^{a}${\it DAMTP, University of Cambridge,} }
\centerline{\it Wilberforce Road, Cambridge CB3 0WA - UK. }
\medskip
\centerline{${}^{b}${\it Department of Physics,} }
\centerline{{\it University of Wales Swansea,} }
\centerline{\it Singleton Park, Swansea SA2 8PP - UK.}

\bigskip

\noindent

We obtain the super-Landau-Ginzburg mirror of the A-twisted
topological sigma model on a twistor superspace -- the quadric
in $\mCP^{3|3}\times \mCP^{3|3}$ which is a Calabi-Yau
supermanifold. We show that the B-model mirror has a geometric
interpretation. In a particular limit for one of the K{\"a}hler
parameters of the quadric, we show that the mirror can be interpreted
as the twistor superspace $\mCP^{3|4}$. This agrees with the
recent conjecture of Neitzke and Vafa proposing a mirror equivalence
between the two twistor superspaces.

\noindent

\Date{2004}

\lref\Witten{E. Witten, {\it Perturbative gauge theory as a String
Theory in Twistor Space}, hep-th/0312171 }

\lref\Neitzke{
A.~Neitzke and C.~Vafa,
``N = 2 strings and the twistorial Calabi-Yau,''
arXiv:hep-th/0402128.}

\lref\AganagicYH{M.~Aganagic and C.~Vafa,
``Mirror symmetry and supermanifolds,''
arXiv:hep-th/0403192.}

\lref\Wittenold{E.~Witten,
``An Interpretation Of Classical Yang-Mills Theory,''
Phys.\ Lett.\ B {\bf 77}, 394 (1978).}

\lref\NekrasovJS{
N.~Nekrasov, H.~Ooguri and C.~Vafa,
``S-duality and topological strings,''
arXiv:hep-th/0403167.}

\lref\Kapustin{
A.~Kapustin,
``Gauge theory, topological strings, and S-duality,''
arXiv:hep-th/0404041.}

\lref\Rocek{
M.~Rocek and E.~Verlinde,
``Duality, quotients, and currents,''
Nucl.\ Phys.\ B {\bf 373}, 630 (1992)
[arXiv:hep-th/9110053].}


\lref\schwarz{A.~Schwarz,
``Sigma models having supermanifolds as target spaces,''
Lett.\ Math.\ Phys.\  {\bf 38}, 91 (1996)
[arXiv:hep-th/9506070].}

\lref\phases{
E.~Witten,
``Phases of N = 2 theories in two dimensions,''
Nucl.\ Phys.\ B {\bf 403}, 159 (1993)
[arXiv:hep-th/9301042].}

\lref\HoriKT{
K.~Hori and C.~Vafa,
``Mirror symmetry,'' arXiv:hep-th/0002222.}

\lref\SethiCH{
S.~Sethi,
``Supermanifolds, rigid manifolds and mirror symmetry,''
Nucl.\ Phys.\ B {\bf 430}, 31 (1994)
[arXiv:hep-th/9404186].}

\lref\horivafa{
K.~Hori, A.~Iqbal and C.~Vafa,
``D-branes and mirror symmetry,''
arXiv:hep-th/0005247.}

\lref\AtiyahWI{
M.~F.~Atiyah, N.~J.~Hitchin and I.~M.~Singer,
``Selfduality In Four-Dimensional Riemannian Geometry,''
Proc.\ Roy.\ Soc.\ Lond.\ A {\bf 362}, 425 (1978).
}

\lref\WardVS{
R.~S.~Ward and R.~O.~Wells,
``Twistor Geometry And Field Theory,'', Cambridge, UK: Univ. Pr. (1990). 
}

\newsec{Introduction}

Recently Witten \Witten\ has argued that perturbative ${\cal N}=4$
supersymmetric $U(N)$
Yang-Mills theory can be formulated as topological string theory
with the supertwistor space ${\mCP}^{3|4}$ as target. Among other
things this has led to new and interesting observations about certain
Calabi-Yau supermanifolds \Neitzke\AganagicYH\ which happen to be
supertwistor spaces and these form the focus of our note.

In particular, in \Witten\ it was demonstrated that the  ${\cal N}=4$
Yang-Mills
amplitudes, when transformed to the supertwistor space ${\mCP}^{3|4}$, are
supported on holomorphic curves which were then interpreted as
D1-instantons of the topological B-model on ${\mCP}^{3|4}$. On the
other hand it was shown long ago
\Wittenold\ that the the classical equations of motion of the
${\cal N}=4$ gauge theory follow from integrability of gauge fields on
supersymmetric lightlike lines. The space of all such lightlike lines
in (complexified) compactified Minkowski space is the quadric in
${\mCP}^{3|3}\times{\mCP}^{3|3}$. It is natural to ask if there is
some relation between these two pictures. In \Neitzke\ it was
conjectured that these two twistorial formulations could possibly be
related by mirror symmetry between ${\mCP}^{3|4}$ and the quadric in
${\mCP}^{3|3}\times {\mCP}^{3|3}$. (They are both Calabi-Yau
supermanifolds). This conjecture 
was prompted by a combination of two observations. First, the authors
of \Neitzke\ argued that the ${\cal N}=4$ Yang-Mills amplitudes could
also be obtained from the A-model topological string on
${\mCP}^{3|4}$ to be understood as $S$-dual (see also \NekrasovJS\ and
\Kapustin) to the B-model
picture of \Witten. In this picture the D1-instantons of \Witten\ are
replaced by worldsheet instantons of the A-model (and the D5-branes by
NS5-branes). Secondly, given such an
A-model description, it is natural to expect that a potential mirror B-model
description will have no instantons and the perturbative Yang-Mills
amplitudes will
be realized classically. The observations of \Wittenold\ outlined
above suggest a candidate for such a mirror. Specifically,
one expects that the B(A)-model on ${\mCP}^{3|4}$ is mapped by a
mirror transformation to the A(B)-model on the quadric in ${\mCP}^{3|3}
\times{\mCP}^{3|3}$. In
a recent work \AganagicYH\ it was demonstrated that the
A-model on the former supermanifold is mirror to the B-model on the
latter in the limit where the K\"ahler class $t$ of $\mCP^{3|4}$ is
sent to minus infinity.

In this note we show that the A-model on the quadric in ${\mCP}^{3|3}
\times{\mCP}^{3|3}$ is mirror to the B-model on ${\mCP}^{3|4}$ in
a certain limit for one of the two K{\"a}hler parameters of the
quadric. The motivation for this study is two-fold. One would like to
understand if indeed the two twistorial formulations above are
related by mirror symmetry, independently of the conjectured
$S$-duality for topological strings. Further, one would like to shed
light on the relationship between K{\"a}hler class and
complex structure deformations of the two supermanifolds in question.
It should be pointed out that we are studying the closed topological
model (without any extra branes) which corresponds to the
gravitational theory. Our results show that the B-model mirror of the
quadric should be understood as a complex deformation of the
twistor superspace ${\mCP}^{3|4}$ (with a line at infinity removed). It has 
been argued that complex deformations of twistor space get mapped to
$\mC^4$ (complexified Minkowki space) with points blown up (see
\NekrasovJS\ and references therein). It would be extremely
interesting to develop this idea further.

In the following section we review some essential results in the context of
mirror symmetry for supermanifolds. In Section 3 we apply these to the 
A-model on the quadric in $\mCP^{3|3}\times\mCP^{3|3}$ and obtain the
mirror B-model which has a geometric interpretation. 
 
\newsec{Supermanifolds and Hypersurfaces in Toric Manifolds}

We begin by reviewing the results of \schwarz\ and \AganagicYH\ which are
relevant for our computation. We are interested in computing the
mirror transform of the topological sigma model of the A-type on the
quadric \foot{For the sake of brevity we will refer to the quadric
in ${\mCP}^{3|3}\times{\mCP}^{3|3}$ simply as ``the quadric''.}
which is realized as a hypersurface in a toric
supermanifold.
\subsec{Degree $d$ hypersurface in ${\mCP}^{d-1}$}
To understand how this proceeds we
first recall the well-known fact \HoriKT\ that the
observables of the A-model on a bosonic Calabi-Yau manifold ${\cal M}$
realized
as a hypersurface in a compact toric manifold are simply related to
the observables of the A-model on a corresponding {\it non-compact} toric
manifold $V$.
As a simple example, consider a hypersurface obtained
from a degree $d$ polynomial equation in ${\mCP}^{d-1}$. This is
realised as a $U(1)$-gauged linear sigma model with ${\cal N}=(2,2)$
supersymmetry and
$d$ chiral superfields $\{\Phi_i\}$ of charge $+1$ each. In addition
there is a field $P$ of charge $-d$ and a superpotential,
\eqn\sup{W=\epsilon \;P \;G(\Phi_i)}
where $G(\Phi_i)$ is a degree $d$ polynomial (weight $d$) \phases. These
lead in the infrared to a non-linear sigma model description with the
Calabi-Yau ${\cal M}$ as target via the vacuum equations
\eqn\nlsm{G(\Phi_i)=0;\quad P=0,}
modulo complex gauge transformations.
However, the observables of the A-model (the ({\it a,c})-ring)
cannot depend on the
superpotential which is a ({\it c,c})-ring deformation. Therefore we
expect that as $\epsilon \rightarrow 0$ and the superpotential
disappears, the observables of the A-model on ${\cal M}$ are simply related
to the observables of the {\it non-compact} Calabi-Yau manifold $V$ which is
the ${\cal O}(-d)$ line bundle over ${\mCP}^{d-1}$ (the field $P$ is
then identified with the coordinate on the fibre). Of course, this
reasoning is not strictly correct since the two theories differ
drastically, as the one with vanishing superpotential 
has 2 extra complex dimensions. Nevertheless, one indeed obtains
the following correspondence between states
of the A-twisted theory on ${\cal M}$ and those on $V$,
\eqn\map{|\Sigma\rangle_V \rightarrow |\bf 1\rangle_{\cal M}}
where $\Sigma$ is the twisted chiral abelian field strength and the
state $|\Sigma\rangle_V$ is a normalizable ground state of the
noncompact theory obtained by an insertion of $\Sigma$ which
corresponds to the K{\"a}hler form
controlling the size of the compact part of the
geometry. Alternatively, such $\Sigma$-insertions
can be achieved by taking derivatives of correlators
of the A-twisted model on $V$ with respect to
the complexified K{\"a}hler parameter $t$, yielding A-model observables on
the compact manifold ${\cal M}$.
\subsec{Realization as the supermanifold ${\mWCP}^{d-1|1}(1,\ldots,1|d)$}

It was shown in \schwarz\ that the above procedure of computing
A-model observables of the compact manifold ${\cal M}$ is equivalent
to A-model computations on a $(d-1|1)$-dimensional toric
{\it supermanifold}. Specifically, in the example above, it simply
amounts to replacing $P$ by a fermionic chiral superfield $\Psi$ of charge
$d$. Thus we have an ${\cal O}(d)$ (fermionic) bundle over
${\mCP}^{d-1}$ which can also be viewed as the toric supermanifold
$\hat {\cal M}$ namely, ${\mWCP}^{d-1|1}(1,\ldots,1|d)$. The D-term
constraint is now,
\eqn\dsup{\sum_{i=1}^{d}|\Phi_i|^2+d\;\bar\Psi\Psi={\rm Re}[t]}
where the K{\"a}hler class parameter of $\hat {\cal M}$ is the same as
that of the compact Calabi-Yau ${\cal M}$. Since $\hat {\cal M}$ has
$U(1)$ isometries one can perform T-duality to obtain the
super-Landau-Ginzburg mirror. Techniques for doing this have been
discussed in \SethiCH\ and \AganagicYH. Interestingly the upshot of
this procedure is that the super-Landau-Ginzburg directly yields the
observables, such as periods of the compact bosonic Calabi-Yau $\cal
M$. Put another way, the fermionic fields of the sigma model on $\hat
{\cal M}$ automatically
incorporate the projection (the $t$-derivative)
that was required above to translate the observables of the
non-compact manifold $V$ into those of the compact Calabi-Yau ${\cal M}$.
\subsec{Review of T-duality for fermionic coordinates}

The implementation of T-duality for fermionic coordinates has
only recently been discussed in \AganagicYH. Since it is not part of
standard literature we review the main results here which will be used
subsequently. Just as in the case of bosonic coordinates with a $U(1)$ 
isometry
\Rocek\ we wish to dualize the phase for a fermionic superfield
$\Psi$ with a $U(1)$-charge $q$. The phase is bosonic and hence it
will dualize into a bosonic 
twisted chiral multiplet $Y$. The real part of $Y$ is determined as
$Y+\bar Y =\bar \Psi\Psi$ while its imaginary part is periodic. 
In addition, the usual twisted chiral
superpotential is also generated for $Y$ which gives the winding modes
a mass $q\Sigma$. However, this is not all.
The original theory had one fermionic coordinate $\Psi$ whose momentum
modes have mass $q \Sigma$. Hence the
dualized theory cannot simply have one bosonic degree of freedom. In
fact, it should have two fermionic superfields $\eta, \chi$ with the
same mass $q\Sigma$ as the winding modes of the dual bosonic coordinate
$Y$. This ensures that one
boson and one fermion cancel in the partition function. In sum then, the
T-dual of the fermionic superfield $\Psi$ yields the bosonic twisted chiral
multiplet $Y$ and two fermion superfields $\eta,\chi$ with a
superpotential
\eqn\ferdual{W= - q\;\Sigma(Y-\eta\chi) +e^{-Y}}
This superpotential gives the same mass $-d\;\Sigma$ to the winding
modes of $Y$ and the excitations of $\eta, \chi$. We can rewrite this
superpotential in a different form after a shift $Y \rightarrow
Y+\eta\chi$ so that
\eqn\ferdualsup{W=-q\;\Sigma Y + e^{-Y}(1-\eta\chi).}
Now we see what the effect of the fermions is on the partition
function of the dual theory. Integrating them out brings
down a factor of $e^{-Y}$ in the measure turning $e^{-Y}$ into a good
coordinate. In the context of the example discussed above this is precisely
the effect of taking a $t$-derivative of the partition function of the
theory corresponding to the bosonic non-compact manifold $V$.
\newsec{Mirror of the quadric}

We are now ready to apply the results above to the case of interest.
We take two copies of ${\mCP}^{3|3}$ each with the homogeneous
coordinates $\{X_I,\Psi_A\}$ and $\{\tilde X_I,\tilde \Psi_A\}$
respectively where $\{X_I,\tilde X_I\}$, ($I=1,\ldots,4$) are
bosonic coordinates and $\{\Psi_A,\tilde\Psi_A\}$, ($A=1,2,3$)
are the fermionic coordinates. Let $t_1$ and $t_2$ be the complexified
K{\"a}hler class parameters for the two $\mCP^{3|3}$'s.
The quadric $Q$ in ${\mCP}^{3|3}\times
{\mCP}^{3|3}$ is then defined by the bilinear equation
\eqn\defquadric{G:=\sum_{I=1}^4X_I\tilde X_I +\sum_{A=1}^3\Psi_A\tilde\Psi_A=0}
There is a ${\bf Z}_2$ symmetry which exchanges the two
${\mCP}^{3|3}$s and their K{\"a}hler classes $t_1\leftrightarrow t_2$.
This quadric can be realised as a $U(1)\times U(1)$ gauged
linear sigma model with the charge assignments $(1,0)$ for the fields
$\{X_I,\Psi_A\}$ and $(0,1)$ for the second copy of coordinates
$\{\tilde X_I,\tilde \Psi_A\}$. In addition we introduce a
bosonic chiral superfield $P$ of charge $(-1,-1)$ and a superpotential
\eqn\supquadA{W = \epsilon P G[X_I, {\tilde X}_I\Psi_A,\tilde\Psi_A].}
The field content and charge assignments ensure that there is no
$U(1)_A$ anomaly and the Calabi-Yau condition is satisfied. (Note that
due to the reversed statistics
the fermionic coordinates contribute to the anomaly with a sign
opposite to that of the bosonic coordinates.)
The theory flows to a (super)Calabi-Yau phase where it is a non-linear
sigma model with the quadric $Q$ as target, corresponding to the vacuum
manifold
\eqn\vacuum{G[X_I,\tilde X_I,\Psi_A,\tilde\Psi_A]=0}
with $P=0$ in the D-term constraints,
\eqn\dvacone{\eqalign{&\sum_{I=1}^4|X_I|^2+
\sum_{A=1}^3|\Psi_A|^2-|P|^2 = {\rm Re}[t_1]\cr
&\sum_{I=1}^4|\tilde X_I|^2+ \sum_{A=1}^3|\tilde\Psi_A|^2-|P|^2 = {\rm
Re} [t_2]}} 
modulo $U(1)\times U(1)$ gauge transformations (${\rm Re}[t_1],{\rm
Re} [t_2]>0$). 
The quadric is a hypersurface in a $(6|6)$-dimensional toric
supermanifold. The bosonic hypersurface equation \vacuum\ means that the
quadric $Q$ has complex (super)dimension $(5|6)$.

In order to implement the mirror transform we need to realise the
topological A-model observables on $Q$ in terms of a sigma model on a toric
(super)manifold {\it i.e.} one without the superpotential which  imposes
the hypersurface constraint. One is naturally tempted to use the
ideas of \HoriKT\ outlined in the previous section namely, to study the
A-model on the ``non-compact'' Calabi-Yau where we send $W\rightarrow
0$. However, this immediately leads to a puzzle -- the sigma model
with $W=0$ has a $(7|6)$-dimensional target space.
On the other hand the quadric has bosonic minus
fermionic dimension $-1$ and its mirror must naturally have dimension
$(n-1|n)$. The resolution is straightforward and we simply need to 
employ the ideas of \schwarz\ and \AganagicYH\ as explained earlier. We
must not only send $W$ to zero but we must also replace the bosonic field
$P$ with a fermionic chiral superfield $\Psi_P$ with charge $(1,1)$
under the $U(1)\times U(1)$ gauge symmetry.

In summary, the topological A-model on the quadric
$Q$ is equivalent to the A-model on the 
${\cal O}_1(1)\otimes{\cal O}_2(1)$ 
fermionic bundle over $\mCP^{3|3}\times\mCP^{3|3}$ (by ${\cal O}_1(1)$
we mean the pullback of the ${\cal O}(1)$ line bundle on the first
$\mCP^{3|3}$ factor, and similarly for the second).

\subsec{B-model mirror}

The Landau-Ginzburg B-model dual of the above can be obtained as
follows. T-duality replaces each bosonic superfield $X_I$ and $\tilde
X_I$ with the cylinder-valued coordinates $Y_I$ and $\tilde Y_I$
respectively, $(I=1,\ldots,4)$. Further, using the rules for dualizing
the fermionic coordinates where each such field yields a bosonic
coordinate and a pair of fermionic fields, the $\{\Psi_A\}$ and
$\{\tilde\Psi_A\}$ dualize to the set $\{M_A,\eta_A,\chi_A\}$ and
$\{\tilde M_A,\tilde\eta_A,\tilde\chi_A\}$ respectively,
$(A=1,2,3)$. In our notation the $\eta$'s
and $\chi$'s are fermion superfields. Finally, the A-model fermion
$\Psi_P$ with charge $(1,1)$ dualizes to $(Y_P,\eta,\chi)$. The
Landau-Ginzburg mirror of the quadric is given by the path integral
(for the holomorphic sector)
\eqn\mir{\eqalign{
Z =  \int \prod_{I=1}^4[dY_I \;d\tilde Y_I]
&\;\prod_{A=1}^3[dM_A \;d \tilde M_A\;d\eta_A d\chi_A \;
d\tilde\eta_A d \tilde \chi_A]\;dY_P\;d\eta d \chi \cr
&  \delta(\sum Y_I -
\sum M_A - Y_P - t_1) \,\;\delta (\sum \tilde Y_I - \sum \tilde M_A -
Y_P 
- t_2) \cr\cr
\times {\rm exp} \left[ \sum e^{-Y_I} + \sum e^{- \tilde Y_I}
\right.&\left. + \sum
e^{-M_A} (1 + \eta_A \chi_A) + \sum e^{-\tilde M_A}
(1 + \tilde \eta_A \tilde \chi_A)
+ e^{-Y_P} (1+ \eta \chi)
\right]}}
The Landau-Ginzburg model has $13$ bosonic (taking into account the
two delta-function constraints) and $14$ fermionic degrees of
freedom. Note that the ${\bf Z}_2$ exchange symmetry of the quadric is
explicit in the B-model partition function above.
To arrive at a mirror super-Calabi-Yau interpretation for this
Landau-Ginzburg we perform a sequence of manipulations that involve
integrating out some of the fields and successive field
redefinitions.

We first integrate out the fermions
$\tilde \eta_A, \tilde \chi_A, \eta_3, \chi_3, \eta, \chi$,
and solve the delta-function constraints for
$Y_P$ and $M_3$. This breaks the symmetry that exchanges the
two ${\mCP}^{3|3}$'s of the A-model. We will come back to this point
later. At this stage we have the following B-model integral with 4
fermions and 13 bosons
\eqn\mirb{\eqalign{
Z = & \int  \prod_{I=1}^4[dY_I \;d\tilde Y_I]\;dM_1 \;dM_2
\prod _{A=1}^3[d \tilde M_A] \;d\eta_1 d\chi_1\;d\eta_2 d\chi_2 \,
\;e^{M_1 + M_2 - \sum Y_I - \sum \tilde M_A + t_1}
\cr
& \times {\rm exp} \left[  \sum e^{-Y_I} + \sum e^{- \tilde Y_I} + e^{-M_1}
(1 + \eta_1 \chi_1) +  e^{-M_2} (1 + \eta_2 \chi_2) + \sum e^{- \tilde
M_A} + \right. \cr
& \left. +  e^{t_1-t_2} e^{M_1+M_2-\sum Y_I + \sum \tilde Y_I - \sum \tilde
M_A} + e^{t_2} e^{\sum \tilde M_A - \sum \tilde Y_I} \right].
}}
We see that integrating out the fermions leads to non-trivial factors in the
measure. These measure factors turn the fields $e^{-Y_I}$ and
$e^{-\tilde{M}_A}$ which were $\mC^*$-valued, into good coordinates, so
that we can define the new $\mC$-valued fields $y_I=e^{-Y_I}$ and
$\tilde m_A = e^{-\tilde{M}_A}$. In fact it is convenient to make a similar
change of variables for all the bosonic fields:
\eqn\change{m_{1,2}:=e^{-M_{1,2}};\;\;\;\tilde m_A:=e^{-\tilde M_A};\;\;
y_I:=e^{-Y_I};\;\;
\tilde y_I := e^{-\tilde Y_I}.}
In terms of these new fields the Landau-Ginzburg model is
\eqn\mirc{\eqalign{
&Z =  e^{t_1} \, \int \prod_{I=1}^4[dy_I]\;
\prod_{I=1}^4\left[{d \tilde y_I \over \tilde y_I}\right]\;
{dm_1 dm_2 \over m_1^2 m_2^2} \prod_{A=1}^3 [d\tilde m_A]\;d\eta_1
d\chi_1\;d\eta_2
d\chi_2 \,\cr
&  {\rm exp} \left[ \sum_{I=1}^4
y_I
+ \sum_{I=1}^4 \tilde y_I + \sum_{a=1}^2 m_a (1+ \eta_a \chi_a) +
\sum_{A=1}^3 \tilde m_A + e^{t_2} {\prod_I \tilde y_I \over \prod_A
\tilde m_A} + e^{t_1-t_2} {\prod_A \tilde m_A \prod_I y_I \over
m_1m_2\prod_I \tilde y_I } \right] \,.
}}
A further change of variable 
\eqn\new{ \tilde x_A := {\tilde y_A \over \tilde m_A},\;(A=1,2,3);
\;\tilde x_4:= \tilde y_4\; ; \qquad
x_a := {y_a \tilde m_a \over \tilde y_a m_a},\; (a=1,2); \;x_b := {y_b
\over \tilde x_b} \, (b=3,4) }
allows us to bring the Landau-Ginzburg superpotential in
the exponent in \mirc\ to a polynomial form which will lead us to the
interpretation as a super-Calabi-Yau manifold:
\eqn\mirb{\eqalign{
Z  = e^{t_1} \int &\prod_{I=1}^4[dx_I]\; \prod_{I=1}^4
[d \tilde x_I] {dm_1 \over m_1} {dm_2\over m_2}
\prod_{A=1}^3[d \tilde m_A]\;
d\eta_1 d\chi_1 d\eta_2 d\chi_2 \cr
&{\rm exp} \left[ \sum_{A=1} ^3\tilde m_A (1+
\tilde x_A) + \sum_{a=1,2} m_a (1+ x_a \tilde x_a + \eta_a \chi_a) +
+  x_3\tilde x_3 + x_4 \tilde x_4 + \right.\cr 
&\left.+\tilde x_4+e^{t_2} \prod_{I=1}^4 \tilde x_I
+ e^{t_1-t_2}\prod_{I=1}^4 x_I \right] \,.
}}
One can now see that the fields $\tilde m_A$ and $\tilde x_4$
are Lagrange multipliers and their equations of motion set $\tilde x_A
=-1$ and $x_4 = e^{t_2}-1$. It is clear from the measure that all the
variables except $m_{1,2}$ are ``good'' variables. The situation can
be rectified following a procedure that is often useful for getting a 
geometric description from the Landau-Ginzburg B-model mirrors of 
Calabi-Yau manifolds (for
instance see \horivafa). 
We introduce additional fields $(u_a,v_a)$, $(a=1,2)$ 
to absorb the non-trivial measure for $m_a$\foot{The idea is to
make use of the relation $\int du dv e^{uv m}={1\over m}$ for a
suitable choice of contour.}. In the resulting
expression $m_1$ and $m_2$ become Lagrange multipliers enforcing
algebraic constraints. Integrating out the Lagrange multipliers we 
finally arrive at the interesting part of the story
\eqn\mire{\eqalign{
Z = & e^{t_1} \int \prod _{a=1,2}[dx_a du_a dv_a d\eta_a d\chi_a]
\prod_a \delta (u_a v_a + \eta_a 
\chi_a - x_a + 1) \, \delta (e^{t_1} (e^{-t_2} -1) x_1 x_2 -1) \,.
}}
The $\delta$-functions inside the integral contain the information on
the geometry of the mirror manifold. The first thing to note is that
the putative mirror geometry has dimension $(3|4)$ (the six bosonic
coordinates have three delta-functions constraints)
consistent with the conjecture of \Neitzke.

How do we understand and interpret this mirror geometry? One
possibility is to take a limit of the A-model K{\"a}hler
parameters in which a simple description appears. Another possibility
might be to homogenize the algebraic constraints and interpret the
mirror as a complete intersection in projective superspace \foot{Even
though strictly speaking the constraints describe a
non-compact manifold, the original geometry we started from, \ie\ the
quadric, is compact and therefore one expects that what we see in
\mire\ is only an affine coordinate patch inside a compact geometry;
but one needs to check that the measure is consistent with such an
interpretation}. However there is not a unique way to homogenize the
equations, and the most obvious possibilities do not result in a
Calabi-Yau supermanifold. Therefore 
we focus on the first possibility and perform a simple rescaling of the
fields to rewrite the Landau-Ginzburg ``period'' as
\eqn\mirf{Z = \int \prod_{a=1,2}[du_a dv_a dx_a d\eta_a d\chi_a] 
\delta (u_1 v_1 +
\eta_1 \chi_1 - x_1 + \nu ) \,  \delta (u_2 v_2 + \eta_2 \chi_2 - x_2
+ \nu)  \, \delta (x_1 x_2 - \mu) \,,}
where $\mu = e^{-t_1}$ and $\nu = (e^{-t_2} -1)^{1/2}$. 

As pointed out earlier, the ${\bf Z}_2$ symmetry under the exchange of
$t_1$ and $t_2$, corresponding to the exchange of the two
$\mCP^{3|3}$ factors in the A-model, has been broken. 
The Landau-Ginzburg integral \mirf\ is a period integral 
over a supermanifold of
dimension $(3|4)$ which we want to identify. (We point out that our
identification of the Landau-Ginzburg integral as a ``period'' is
purely a formal analogy with the case of bosonic Calabi-Yaus. 
For supermanifolds the integrals above vanish unless there are
suitable insertions of fermionic coordinates. For a discussion of
related issues see \SethiCH.)   
We start by considering the limit $\nu \to 0 \sim t_2 \to 0$. 
Then, solving the constraints for
$x_1, x_2$  
\eqn\mirg{
Z = \int \prod_{a=1,2}du_a dv_a d\eta_a d\chi_a
\delta ([u_1 v_1 + \eta_1 \chi_1] [u_2 v_2 + \eta_2 \chi_2] - \mu),}
we can actually perform the delta-function integral by introducing
additional variable changes as follows. In a patch where $u_1\neq 0$
we can introduce the variables 
$z_2=u_2/u_1$, $z_3 =v_1/u_1$, $z_4=v_2/u_1$, $\psi_1=\eta_1/u_1$, 
$\psi_2=\eta_2/u_1$, $\psi_3=\chi_1/u_1$ and $\psi_4=\chi_2/u_1$.
With these new variables the Landau-Ginzburg path integral is
\eqn\mirh{\eqalign{
Z= & \int {du_1 \over u_1} dz_2 dz_3 dz_4 d\psi_1 d\psi_2 d\psi_3
d\psi_4 \delta (u_1^4 (z_2 - \psi_1 \psi_2) (z_3 z_4 - \psi_3 \psi_4)
- \mu) \cr
\equiv &\int {du_1 \over u_1} \Omega_1 \delta ( A u_1^4 - \mu) \cr
=&{1\over \mu} \int \Omega_1 \,,}}
where we introduce the form $\Omega_1 = dz_2 dz_3 dz_4 d\psi_1
d\psi_2 d\psi_3 d\psi_4$.  Note that various factors of $u_1$ in
the measure, induced by the
above variable change, cancel out precisely because of the presence of
fermionic coordinates. The form $\Omega_1$ is the natural holomorphic form on
$\mCP^{3|4}$, given in affine coordinates $z_2, z_3, z_4,\psi_I$, in a
patch where 
one of the homogeneous coordinates (identified with $u_1$)
is set to 1. We can see that this interpretation is valid by
trying to 
write the integral in a different patch. Then we can solve the
$\delta$-function constraints for $u_2$, and we will define some new
coordinates $\tilde z_i, \tilde \psi_I$ related to $z_i, \psi_I$ in
precisely the
way affine coordinates in different patches of $\mCP^{3|4}$ are
related. This will lead to a holomorphic form $\Omega_2$ in the second
affine patch, where $\Omega_1$ and $\Omega_2$ are the same on the
intersection of the two patches. 

This is a confirmation of the conjecture of \Neitzke\ that $\mCP^{3|4}$
is  
the mirror supermanifold of the quadric $Q$ and that this 
interpretation only emerges in a limit of the K{\"a}hler moduli of the
quadric. The Landau-Ginzburg partition function computed in the 
mirror manifold, in the limit where $t_2 \to 0$, is simply
proportional to $e^{t_1}$. This seems to imply that, up to a
normalization, the periods (in this limit for $t_2$) 
do not depend on the K\"ahler
class $t_1$ of the original manifold. Whereas this is in contrast to
the 
usual situation in mirror symmetry, it is consistent with the
arguments of \SethiCH\ , where supermanifolds were proposed as
candidates for mirrors of rigid Calabi-Yaus. That could only be
possible if in these models the K\"ahler class decouples from the
other observables. This issue deserves further study.

It is also worth pointing out that the discrete symmetry
$(t_1\leftrightarrow t_2)$ which
exchanges the two $\mCP^{3|3}$s of the quadric is not visible in the
mirror geometric description \mirg\ obtained from the Landau-Ginzburg
dual \mir. Of course, obtaining the geometric picture required us to
integrate out certain fields which then broke the $t_1\leftrightarrow t_2$ 
symmetry. Obviously we could follow a different route which would
yield the same geometric mirror but with $t_1$ and $t_2$ interchanged in 
Eq.\mirf. It is tempting to speculate that this breaking of the
$t_1\leftrightarrow t_2$ symmetry is intrinsically related to the way
these spaces are defined as twistor spaces. In particular, the 
bosonic part of $\mCP^{3|3} \times\mCP^{3|3}$, in which the quadric
is embedded, is the space  of all self-dual planes and anti-self-dual
planes in $\mC^4$. On the 
other hand, the bosonic part of $\mCP^{3|4}$ is simply the space  
of self-dual (or anti-self-dual) planes in $\mC^4$ and
thus singles out states of a particular helicity.

Finally, we consider the geometric mirror \mirf\ in the general case $\nu
\neq 0$. One expects that it should be interpreted as a complex
deformation of the twistor superspace. 
While $\mCP^{3|4}$ itself may
not have complex deformations, the twistor superspace $\mPT'$ which
should 
actually
be defined as $\mCP^{3|4}\backslash\mCP^{1|4}$ \Witten\ ,
can have complex deformations. This is well-known in the bosonic case 
\AtiyahWI\WardVS\ .
It is interesting to note that if we ignore the fermions
in the delta-functions in \mirf\ and set $\nu=0$, after a simple
variable change it is possible to interpret the resulting expressions
as an ${\cal O}(1)\oplus{\cal O}(1)$ bundle over $\mCP^1$. Turning on
non-zero $\nu$ would be a deformation of this bundle. It
would be interesting to pursue this interpretion in the presence of
the fermions. 
These issues deserve further attention,
particularly if we would like to interpret \mirf\ (for generic
K\"ahler parameters of the quadric) as a complex
deformation of twistor superspace.

{\bf Acknowledgements}: We thank C. Vafa and M. Mari${\tilde{\rm n}}$o
for reading and commenting on a draft of this paper. We acknowledge PPARC for
financial support. 

\listrefs

\bye